\documentclass[12pt,preprint]{aastex}
\usepackage[utf8]{inputenc}
\usepackage{amsmath,amssymb,graphics,graphicx}

\begin{document}
 \title{\bf \large {CLASSICAL SPACE-TIME AS RYDBERG STATES OF UNDERLYING QUANTUM GEOMETRIES}}
 \author{\textbf{C. Sivaram} \\ Indian Institute of Astrophysics, \\ 
 Sarjapur Road, 2nd Block, Koramangala, Bangalore, 560034, India}
\email{profcsivaram@gmail.com}
 \vspace{4cm}
 {\textit{Essay written for the Gravity Foundation 2016 Award for Essays on Gravitation; \\ Submitted on: March 30, 2016}}

\section*{Summary}
Classical macroscopic space-time is pictured in terms of Rydberg states of an underlying discretized ‘atomic’ quantum geometry at Planck scales. While quantum geometry on such scales involves several very short lived transitions changing curvature and topologies, the Rydberg states have very long lifetimes, going as a high power of the quantum number $n$. This means space-time on macroscopic scales are almost infinitely stable. The large degeneracy in the Rydberg levels, with high $n$, can also account for a large black hole entropy, as well as long lifetime of massive black holes to quantum decays. We have a possible promising paradigm to link quantum geometry at Planck scales, to classical space-time.

Keywords: Black hole entropy, Classical space-time, Correspondence principle, Planck scales, Quantum geometry, Rydberg states    
\newpage

\section{Description}
The recent landmark discovery by the LIGO group [1] of direct detection of gravitational waves emitted by coalescing compact objects, triumphantly supports yet another prediction of Einstein’s general relativistic theory of gravitation. Though a hundred years separates the prediction and the actual discovery, the result is as epochal as Hertz’s verification of Maxwell’s prediction of electromagnetic waves. In the case of electromagnetic waves, this was followed by rapid progress in the next two decades of the understanding of myriad properties of the interaction of electromagnetic field quanta (i.e. photons) with matter, i.e. various types of scattering emission and absorption phenomena (involving photons). This led to the spectacular development in quantum mechanics (the Bohr model, uncertainty principle etc.) and quantum field theory. However in the case of gravity, the weakness of the interaction (some 36 orders weaker than electromagnetism) precludes any corresponding progress in the detection of quantum processes involving emission absorption and scattering of gravitons. At the cross sections are several orders smaller, it may take several decades to obtain direct evidence for gravitons and their interaction with matter! (The $3d\rightarrow1s$ atomic transition involving gravitons has lifetime $\sim 10^{38}$ s [2]). Of course, it is formally possible to estimate the number of gravitons emitted, for example, in the event detected by LIGO, the number of gravitons (of kilohertz frequency) works out to be $\sim 10^{78}$! In an asymmetric explosion of energy $E$, the number of gravitons is typically [3, 4]
\begin{equation}
 N\sim \frac{G}{c^5}\frac{E^2}{\hbar}
\end{equation}
For two orbiting bodies (masses $m$ and $M$), the number of gravitons in an orbital period is [4],
\begin{equation}
N\sim \frac{64 \pi G^3}{5c^5\hbar} \frac{m^2 M^2}{R^2}.
\end{equation}
This is about $\sim 10^{54}$ for the Sun- Jupiter system. The cross section for interaction with matter is however incredibly low, despite the large numbers. But the basic difference between gravity and electromagnetism is the rapid rise in cross sections (and rate of interaction) with energy of the former. For instance the cross sections for annihilation of electron – positron pairs to gravitons, rises with energy $E$ as $\sim G^n E^{2n}$, while for electromagnetism (i.e. conversion to photons) it is only $\sim ~ \ln E$. At Planck energies, $E\sim 1/G~ (\hbar=c=1)$, the processes become stronger than for electromagnetism, (or any gauge vector fields) and graviton emission (or scattering) is dominant [5, 6]. The timescales are $\sim 10^{-43}$ s. In the collapse of the matter to such super high densities (or temperatures) in black holes (interiors) or in the early universe, such processes would be important. For example, in the case of the Bohr model of atoms, it was realized that the accelerating electrons soon radiate all their energy and collapse into the nucleus, i.e. collapse into a singularity, making all atomic configurations unstable! This classical collapse (to a singularity) by emission of electromagnetic waves was averted by quantum theory postulates, that in stationary orbits (with angular momenta, in multiples of $\hbar$), the electrons would not radiate and atoms would be stable. Similar postulates (like the Bohr-Sommerfeld rule) would prevent gravitational singular collapse. Moreover space-time at quantum (Planck) scales would be unstable (all Planck mass entities etc.) and far from smooth stable classical (macroscopic) space-time. This also applies to superstring or loops, etc. While we can consider all particles to be generated by vibration of strings, those entities, themselves would decay (for example, under their high tension, $\sim \alpha c^2/G$) [7, 8].
A Bohr-Sommerfeld system of Planck entities (atoms) could serve as a model for quantum geometry. Their separations would be multiple of $L_{pl}$ (stationary orbits defined in the usual way or in terms of phase space being multiples of $\hbar^3$ would prevent decay and render stability). In the case of strings, the equivalent rule would be [9]
\begin{equation}
\frac{1}{c} \int T \mathrm{d}A = n \hbar.
\end{equation}
Here $T$ is the tension, and $\mathrm{d}A$ the element of area swept. At the Planck scale underlying the level of quantum geometry, we can talk in terms of areas and curvatures. The scale corresponds to curvature $K_{pl}\sim c^3/\hbar G$. This upper bound on the curvature is finite but very large. This is analogous to the $H$ atom where energy is unbounded from below classically, but because of $\hbar$, has a finite value $\displaystyle{E_v= - \frac{m e^4}{n^2}\frac{1}{\hbar^2}}$. As $\hbar \rightarrow \infty$, here $E_0$ is unbounded, and so also the curvature. A self adjoint curvature operator corresponding to curvature can be constructed from quantum geometry and is bounded from above. So at Planck scales, quantum geometry replaces classical differential geometry. Indeed in models like loop quantum gravity (LQG), the discreteness of quantum geometry [10, 11] is embodied by having quantized eigenvalues of the area operator (associated with a 2-surface) given by
\begin{equation}
 A_j \propto 8 \pi L_{pl}^2 \sum_i \sqrt{J_i (J_i +1)}.
\end{equation}
After all gravity is space-time itself and any deep-rooted quantum approach should involve quantized geometry parameters like area or volume. Just as we have quantized energy eigenvalues in atomic spectroscopy (like in the Bohr or the Dirac model), here we have a spectrum of quantized areas (or curvature which has the dimension of the inverse area) or volumes. The operator eigenvalues crowd rapidly, as areas and volumes increase (they are not uniformly spaced). We shall discuss a continuum approximation which results in classical space-time. Now at Planck scales, we can picture the quantum geometry undergoing numerous transitions (energy levels) with discrete (continuous) changes in the curvature (or areas). There are spontaneous disintegration and transitions going on all the time in quantum space-time foam. But what makes classical space-time stable? How can we understand the stability of space-time at macroscopic scales? Our picture here has some similarities to LQG. We have for quantum geometry, basically a Bohrian atomic picture (where though the entities are stable and do not decay due to analogous quantization rules) there are transitions (with short lifetimes) giving rise to changes in topology. Energy densities change, giving rise to curvature transitions. In an analogous string version, we had relations like [12, 13]
\begin{equation}
 string~ tension \times curvature \sim \hbar.
\end{equation}
Also, the associated phase space (both for black hole interiors, strings and other entities) is always $\hbar^3$.
Now in the Bohr atomic picture, we have the so-called correspondence principle, i.e. the frequency given by quantum theory for two very large quantum numbers and separated by unity, becomes identical with classical (orbital) frequency. Thus, if $n_1$ and $n_2$ are relatively high quantum numbers
\begin{equation}
\nu = R c \left(\frac{1}{n^2}-\frac{1}{(n+\Delta n)^2}\right)\simeq R c \frac{2 \Delta n}{n^3},
\end{equation}
where $R$ is the Rydberg constant. If $\Delta n=1$, $\displaystyle{\nu=\frac{2 R c}{n^3}}$, which is the classical frequency. The so-called Rydberg states correspond to very high values of $n$ ($n$ around 1000 have been observed in astrophysics). The Rydberg atoms have large size scaling as:
\begin{equation}
 d= n^2 a_B,
\end{equation}
 where $a_B$ is the Bohr radius. Here, for quantum geometry (made up of appropriate ‘atoms’ of Planck entities), we have $a_B\simeq L_{pl}$. Thus $n\approx 10^{10}$, would give nuclear radii, etc. So in this viewpoint, classical space-time would correspond to a high value of n. Now the decay rates of Rydberg states, is given by [14]: 
\begin{equation}
 \Gamma = \frac{d^2 \omega^3}{\hbar c^3}.
\end{equation}
The matrix element for electric dipole transition, between states of $n$, $n’$, for $\Delta n= n'-n=1$ is given by
$ d=\langle m|\hat{d}|n'\rangle \sim qn^2a_B$. Using this in eq. (8), we get
\begin{equation}
 d^2 \propto n^4, \quad \omega^3 \propto n^{-9}.
\end{equation}
So the transition rates scales with $n$, as $\Gamma=\Gamma_0 n^{-5}$, where $\Gamma_0=c\alpha^4/a_B$, so 
that the lifetime of the state is given by
\begin{equation}
 t=t_0 n^5, 
\end{equation}
with $t_0=10^{-9}$ s. So for a high $n$, one can have very long lifetimes. We have an analogous situation now for quantum geometry. The Rydberg states correspond to classical (macroscopic) geometry. The decay timescales now depend on the quadrupolar matrix element $\langle n|\ddot{\hat{Q}}|n' \rangle$; with $\ddot{\hat{Q}}\sim d^4 \omega^6$, the scaling now going as $n^8$. Thus, the lifetime of the state is 
\begin{equation}
 t=t_{pl} n^8,
\end{equation}
with $t_{pl}$ being the Planck timescale. (Above, for the electromagnetic case we had, $t_0$ connected to $\alpha$, the fine structure constant). Thus, with $n\approx 10^{10}$ (characterizing the nuclear length scale), the space-time would be stable for $10^{37}$s! Classical space-time in this picture corresponds to Rydberg states (with high values of $n$) of the underlying quantum geometry, while transitions involving the quantum geometry have short (Planck or multiples of Planck) timescales. The classical geometry corresponding to Rydberg states (or levels) of this quantum geometry have very long lifetimes and are stable. At macroscopic scales, we have arbitrarily long lifetimes to decay or transitions. Large curvatures (small areas) decay fast while small curvatures (large areas) have long lifetimes. There is some analogy to LQG, where for large eigenvalues (eg. See eq.(3)), the separation $\Delta a_n=a_{n+1}-a_n$ between consecutive eigenvalues, decreases exponentially (here we have a power law decrease). Due to such strong crowding, the continuum approximation becomes appropriate even a few orders of magnitude above Planck scale. We have this alternate picture, where quantum (discrete) geometry goes over into classical continuum geometry for large values of $n$. As seen above, the curvature scales as $1/n^4$, and so also the energy densities, which is what is expected in GR. The areas are quantized as $(n^2L_{pl})^2$. High $n$ corresponds to large area and low curvature and long lifetimes (which scale as $n^8$, see eq. (12)). So although the ‘atomic’ structure of quantum geometry includes transitions (energy and curvature changes) of short timescales, classical macroscopic space-time is characterized by almost infinitely long lifetime Rydberg-type transitions. There is an analogy of correspondence principle here, connecting quantum and (macroscopic) classical space-time for large values of $n$. ($n=10^{10}$ gives the forty order decrease in the curvature and a $10^{80}$ increase in decay times as compared to $t_{pl}$). We can also apply this picture to understand black hole entropy. For a solar mass black hole, the horizon radius $\approx 10^{38} L_{pl}$, so we require $n\approx 10^{19}$. The degeneracy is seen to be $\sim n^4$. So we have a factor of $10^{76}$ increase in the number of micro-states. If each of the states has $k_B$ entropy, we get $\approx 10^{76} k_B$, which is indeed the entropy of such a black hole. It has been noted before [15], that the entropy scales inversely as the energy density at the black hole horizon (as compared to Planck density). Now density at the horizon is
\begin{equation}
 \rho_h\approx\frac{c^6}{G^3 M^2}.
\end{equation}
So entropy scales as $M^2$, or as compared to $M_{pl}$, it is
\begin{equation}
 S=k_B \left(\frac{M}{M_{pl}}\right)^2.
\end{equation}
The stability of the classical space-time (with large value of n) ensures long lifetime for massive black holes against quantum decay (going as $n^8$), $\sim 10^{80}$ s for a solar mass black hole. Future work to elaborate on this new picture [16] would include setting up of an appropriate Schroedinger-Dirac equation for the quantum geometry, spectral solutions and connection with classical probabilities through an analogous Fokker-Planck equation. Some aspects of this has already been applied to the early universe [17].

\section*{References}
\begin{enumerate}
 \item Abbott, B.~P., Abbott, R., Abbott, T.~D., et al.\ 2016, Physical Review Letters, 116, 061102 
 \item Weinberg, S.\ 1972, Gravitation and Cosmology: Principles and Applications of the General Theory of Relativity, by Steven Weinberg, pp.~688.~ISBN 0-471-92567-5.~Wiley-VCH , July 1972., 688 
 \item Misner, C.~W., Thorne, K.~S., \& Wheeler, J.~A.\ 1973, San Francisco: W.H.~Freeman and Co., 1973
 \item Sivaram, C., CURRENT SCIENCE-BANGALORE- 79.4, 2000, 413-420.
 \item Sivaram, C., \& Sinha, K.~P.\ 1979, Phys. Reports, 51, 111, 
 Susskind, L.\ 1995, Journal of Mathematical Physics, 36, 6377
 \item Sivaram, C.\ 2015, International Journal of Modern Physics D, 24, 1544023 
 \item Sivaram, C.\ 1987, Nature, 327, 108 
 \item Susskind, L., \& Lindesay, J.\ 2005, An introduction to black holes, information, and the string theory revolution : the holographic universe / Leonard Susskind, James Lindesay.~ Hackensack, NJ : World Scientific, c2005,  
 \item Sivaram, C.\ 1994, International Journal of Theoretical Physics, 33, 2407 
 \item Rovelli, C.\ 2004, Quantum Gravity, by Carlo Rovelli, pp.~480.~ISBN 0521837332.~Cambridge, UK: Cambridge University Press, November 2004., 480 
 \item Gambini, R., \& Pullin, J.\ 2014, Classical and Quantum Gravity, 31, 115003 
 \item Corda, C., Hendi, S.~H., Katebi, R., \& Schmidt, N.~O.\ 2013, Journal of High Energy Physics, 6, 8 
 \item Salam, A., \& Strathdee, J.\ 1978, PRD, 18, 4480 
 \item Herzberg, G.\ 1945, New York: Dover, 1945,  
 \item Sivaram, C.\ 2001, General Relativity and Gravitation, 33, 175 
 \item Sivaram, C.\ 2016, work in progress
 \item Sivaram, C., \& Arun, K.\ 2012, Astrophysics and Space Science, 337, 767 
\end{enumerate}

\end{document}